\documentclass{elsart}

\usepackage{graphicx}
\usepackage{amssymb}

\begin{document}

\newcommand{\T}[1]{(TMTSF)$_2$#1}       
\newcommand{\PF}{\T{PF$_6$}}
\newcommand{\X}{\T{X}}

\newcommand{\mTst}{T^\star}         \newcommand{\Tst}{$\mTst$}
\newcommand{\mTsdw}{T_{\mathrm{SDW}}}       \newcommand{\Tsdw}{$\mTsdw$}

\newcommand{\mv}[1]{\mathbf{#1}}                
\newcommand{\vek}[1]{$\mv{#1}$}                 
\newcommand{\parl}[2]{\mv{#1}\|\mv{#2}}         

\newcommand{\mva}{\mv{a}}   \newcommand{\va}{$\mva$}
\newcommand{\mvb}{\mv{b'}}  \newcommand{\vb}{$\mvb$}
\newcommand{\mvc}{\mv{c^*}} \newcommand{\vc}{$\mvc$}

\newcommand{\mBc}{\parl{B}{c^\star}}    \newcommand{\Bc}{$\mBc$}

\newcommand{\plane}[2]{#1\mbox{--}#2}           
\newcommand{\macplane}{\plane{\mva}{\mvc}}      \newcommand{\acplane}{$\macplane$}
\newcommand{\mbcplane}{\plane{\mvb}{\mvc}}      \newcommand{\bcplane}{$\mbcplane$}
\newcommand{\mabplane}{\plane{\mva}{\mvb}}      \newcommand{\abplane}{$\mabplane$}
\newcommand{\abplaneo}{$\plane{\mva}{\mv{b}}$}

\newcommand{\veps}{\varepsilon_0}

\begin{frontmatter}
\title{Hall resistivity in unconventional spin density wave in \PF\ below $T=4.2\,$K}

\author[PMF]{M. Basleti\'{c}\corauthref{email}},
\corauth[email]{Corresponding author} \ead{basletic@phy.hr}
\author[IFS]{B. Korin-Hamzi\'{c}},
\author[PMF]{A. Hamzi\'{c}},
\author[USCLA]{K. Maki}

\address[PMF]{Department of Physics, Faculty of Science, POB 331, HR-10001 Zagreb, Croatia}
\address[IFS]{Institute of Physics, POB 304, HR-10001 Zagreb, Croatia}
\address[USCLA]{Department of Physics, University of Southern California, Los Angeles,\\ CA 90089-0484, USA}

\begin{abstract}
It is well documented that SDW in \PF\ undergoes another phase transition at $\mTst \approx 4\,$K, though the
nature of the new low temperature phase is controversial. We have shown recently that the new phase is well
described in terms of unconventional SDW (USDW) which modifies the quasiparticle spectrum dramatically. In
this paper we show that the same model describes consistently the Hall resistivity observed in \PF.
\end{abstract}

\end{frontmatter}

\section*{Introduction}

Since the discovery of superconductivity in \PF\ in 1979 \cite{JeromeJP80}, the Bechgaard salts or the highly
anisotropic organic superconductors \X\ (where TMTSF is tetramethyltetraselenfulvalene and X is anion PF$_4$,
AsF$_4$, ClO$_4$ \ldots) are one of the most well studied systems \cite{IshiguroBook}. The
quasi-one-di\-men\-si\-o\-na\-li\-ty (1D) is a consequence of the crystal structure, where the TMTSF
molecules are stacked in columns in the \va\ direction (along which the highest conductivity occurs), and the
resulting anisotropy in conductivity is commonly taken to be $\sigma_a : \sigma_b : \sigma_c \approx
10^5:10^3:1$. The rich phase diagram of \X\ salts exhibits various low temperature phases under pressure
and/or in magnetic field, among which the spin density wave (SDW), field induced SDW (FISDW) with quantum
Hall effect and spin triplet superconductivity are very intriguing \cite{LeePRB02}.

\PF\ is metallic down to $\mTsdw\approx 12\,$K, where the transition into the semiconducting SDW takes place.
It is known that SDW in \PF\ undergoes another transition at $\mTst \approx \mTsdw/3$ (at 3.5--4$\,$K at
ambient pressure) \cite{TakahashiJPSJ86,TakahashiSM91,LasjauniasPRL94}. The indication of the subphase was
first seen by NMR \cite{TakahashiJPSJ86}, where $T_1^{-1}$ diverges and the spin susceptibility changes at
\Tst. The transition at \Tst\ is preserved through the entire $p-T$ phase diagram. Furthermore a calorimetric
transition at 3.5$\,$K with a large hysteretic phenomenons in the temperature range 2.5--4$\,$K (caused by
the sample history) has been observed and interpreted as an indication of a glass transition
\cite{LasjauniasPRL94}. On the other hand, the low frequency dielectric relaxation of SDW in \PF\ did not
show the existence of the glass transition \cite{TomicSM97}. Since then the SDW state was widely
investigated, but the nature of the subphase remained controversial. Recently we have studied the \vb\ axis
magnetoresistance (MR) of \PF\ at ambient pressure and with magnetic field rotated within the \acplane\
plane. The MR has different behaviour for $T>4\,$K and $T<4\,$K \cite{KHamzicEPL98}. For $T>4\,$K MR is
described in terms of the quasiparticle in a magnetic field, where the imperfect nesting term
\cite{YamajiJPSJ82,HuangPRB89} plays the crucial role. However, in order to describe MR below 4$\,$K we have
introduced a rather artificial scattering term.

More recently, unconventional density waves (UCDW or USDW) have been proposed as a possible ground state in
electronic systems in organic conductors and heavy fermions \cite{MakiCM0306567}. Unlike the conventional DW,
the UDW is defined as the DW where the order parameter $\Delta(\mv{k})$ depends on the quasi-particle
momentum $\mv{k}$. In particular, UCDW appears to describe the striking angular dependent magnetoresistance
(ADMR) found in the low temperature phase of $\alpha-$(ET)$_2$KHg(SCN)$_4$ \cite{DoraEPL02,MakiPRL03}. On the
other hand, we have shown that the remarkable features of ADMR in \PF\ below $T=4\,$K (the decrease in the
quasiparticle energy gap for $B=0$ and the sudden change in the angular dependence of the energy gap in the
presence of magnetic field $B$ as the temperature crosses $T=\mTst$) can be described within the model SDW
plus USDW using the USDW order parameter $\Delta_1(\mv{k})=\Delta_1\cos2\phi$, where $\phi=bk_2$ with
$\mv{Q}=(2k_F,\pi/2b,0)$ \cite{KHamzicIJMPB02,BasleticPRB02}.

In this paper we shall present the Hall resistivity data in \PF\ for $T>4\,$K and $T<4\,$K and discuss them
within the model of SDW+USDW.

\section*{Hall resistivity}

\begin{figure}
\includegraphics*[width=9cm]{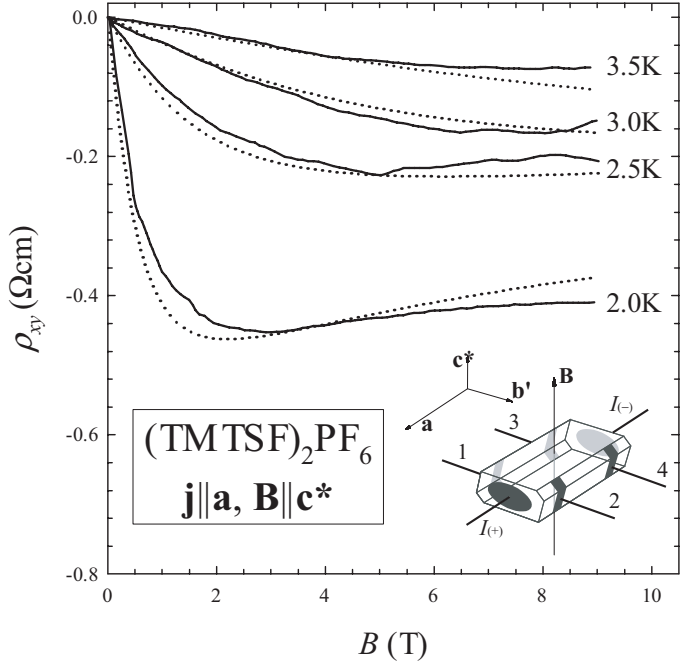}
\caption{Hall resistivity $\rho_{xy}$ versus magnetic field $B$ at several fixed temperatures ($T<4\,$K).
Inset: the contacts arrangement.} \label{fig.fig1}
\end{figure}

The Hall resistivity $\rho_{xy}$ in two crystals of \PF\ with dimension $3.51\times0.61\times0.28\,$mm$^3$
and $3.53\times0.54\times0.25\,$mm$^3$ was measured with 6 contact method as shown in inset of
Fig.\ref{fig.fig1}. The results shown and discussed here were obtained on one of them, and similar
qualitative behaviour was observed on another sample, too. The measurements were performed between 2.0$\,$K
and 6.3$\,$K, with magnetic field up to 9$\,$T. The \va\ direction of the monocrystal is the highest
conductivity direction, the intermediate conductivity \vb\ is perpendicular to \va\ in the \abplaneo\ plane
and the lowest conductivity \vc\ direction is perpendicular to the \abplaneo\ (and \abplane) plane.
The current flow was along the \va\ axis, the magnetic field $\mv{B}$ along \vc\ direction and the Hall
voltage was detected along the \vb\ axis.

\begin{figure}
\includegraphics*[width=9cm]{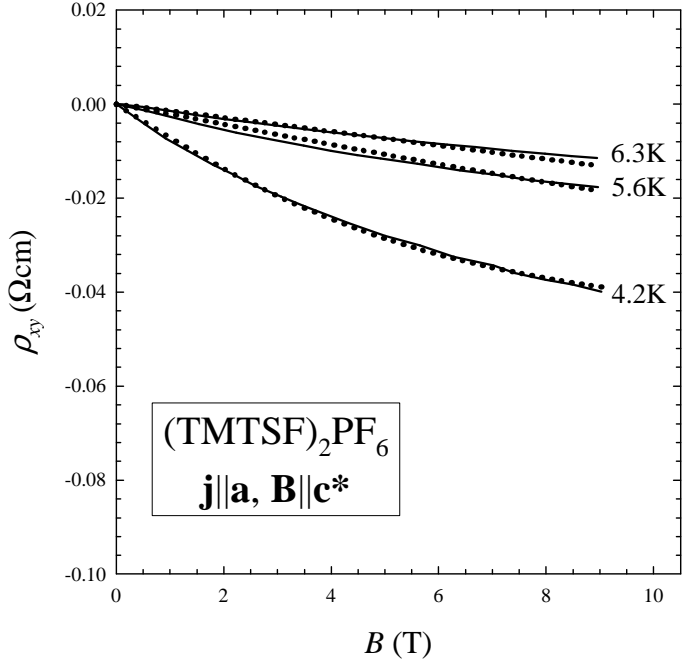}
\caption{Hall resistivity $\rho_{xy}$ versus magnetic field $B$ at several fixed temperatures ($T>4\,$K).}
\label{fig.fig2}
\end{figure}

The magnetic field dependence of the Hall resistivity is shown in Fig.\ref{fig.fig1} ($T<4\,$K) and
Fig.\ref{fig.fig2} ($T>4\,$K).
The general $T$ and $B$ dependencies are consistent with earlier data by Uji {\it et~al.} \cite{UjiPRB97},
although their main objective is the study of the rapid quantum oscillation. As is readily seen from
Fig.\ref{fig.fig1} and \ref{fig.fig2}, the negative Hall resistivity is much smaller for $T>4\,$K than for
$T<4\,$K. According to \cite{MakiPRB90}, the Hall resistivity in the quasi one dimensional system is given by
\begin{equation}
    \rho_{xy}(B) = \frac{\sigma_{yx}}{\sigma_{xx}\sigma_{yy}+(\sigma_{yx})^2}
\end{equation}
where $\sigma_{ij}$ is the conductivity tensor.

If we neglect the quantum Hall effect, which contributes a new term in $\sigma_{xy}$ \cite{YakovenkoJP96}, we
obtain
\begin{equation}
    \sigma_{ij}\sim N_{\mathrm{qp}}\sim e^{-\beta E(B)}\,,
\end{equation}
where $N_{\mathrm{qp}}$ is the quasiparticle density and $E(B)$ is the quasiparticle energy gap in the
presence of magnetic field $B$.

On the other hand, it is well known that $\rho_{xx}(B)$ has no activation form where $x$ is parallel to the
\vek{a}\ axis \cite{IshiguroBook,BasleticPRB02}. A possible explanation is that the conductivity parallel to
the \vek{a}\ axis has another channel of which quasiparticle has no energy gap. A similar approach has been
used in the quantitative analysis of $\rho_{ij}(B)$ in the FISDW state in \PF\ under high pressure
\cite{VuleticEPJB01}. Therefore, the fitting of our Hall data is done with
\begin{equation}
    \rho_{xy}(B) = \frac{A B e^{\beta E(B)}}{1 + C B^n e^{\beta E(B)}}
\end{equation}
where $A$ and $C$ are temperature dependent constants, and
\begin{equation}
E(B)=\left\{     \begin{array}{ll}
        21\,\textrm{K}\times(1+a_>|B|)\,, & \quad \textrm{for } T>\mTst \\
        20\,\textrm{K}\times\sqrt{1+a_<|B|}\,, & \quad\textrm{for } T<\mTst
    \end{array}
\right.
\end{equation}
We took these expressions from \cite{BasleticPRB02}, with $a_>=0.048\,\textrm{T}^{-1}$ and
$a_<=0.027\,\textrm{T}^{-1}$. Here, we limit ourselves to the case \Bc. The results of the fitting procedure
are shown on both figures as the dotted lines (the values of $A$ and $C$ are given in Table \ref{table}). We
have an excellent agreement with the experimental data. We note that $A$ is almost independent of
temperature, while $C$ decreases as temperature decreases. The values of $a$'s used in the present fitting
($a_<=0.011\,\textrm{T}^{-1}$ and $a_>=0.010\,\textrm{T}^{-1}$) are somewhat smaller than the ones used
earlier \cite{BasleticPRB02}, but they are of the same order of magnitude. Also, the exponent $n=1.3$ is
somewhat strange (naturally, we expect $n=2$), but the similar exponent has been found in fitting the
diagonal component of the magnetoresistance tensor \cite{BasleticPRB02}.
\begin{table}
\caption{Table with fitting constants.}\label{table}
\begin{tabular}{c|c|c}
$T\,$(K)        &       $A\,$($\Omega\,$cm/T)        &       $C\,$(T)        \\
\hline   6.3 & $4.9\times10^{-5}$ & $4.0\times10^{-4}$ \\
         5.6 & $4.9\times10^{-5}$ & $4.0\times10^{-4}$ \\
         4.2 & $4.9\times10^{-5}$ & $4.0\times10^{-4}$ \\
         3.5 & $4.9\times10^{-5}$ & $1.0\times10^{-4}$ \\
         3.0 & $4.9\times10^{-5}$ & $1.0\times10^{-4}$ \\
         2.5 & $4.9\times10^{-5}$ & $1.0\times10^{-4}$ \\
         2.0 & $4.0\times10^{-5}$ & $5.4\times10^{-5}$ \\
\end{tabular}
\end{table}
The agreement between the model and experimental data implies that the appearance of USDW below $T=\mTst$,
over the preexisting SDW, with a new quasiparticle energy gap appears to describe the Hall resistivity
consistently. In particular, the rapid increase of the Hall resistivity below $T=\mTst$ testifies the rapid
change in the quasiparticle energy gap across $T=\mTst$. In order to further test the present model the Hall
resistivity data with the magnetic field away from the \vc\ axis are highly desirable.

\section*{Conclusion}

We have completed the study of the resistivity tensor in \PF\ below $T=\mTst$. For \Bc\ we have shown that an
approach with USDW+SDW below \Tst\ gives an excellent fit of the Hall resistivity data. This further supports
our proposal that USDW appearing on top of existing SDW in \PF\ below $T=\mTst$ gives a consistent
description of the resistivity tensor.

\section*{Acknowledgments}

This experimental work was performed on samples prepared by K. Bechgaard. We acknowledge the participation of
N. Franceti\'{c} in the experiment and useful discussions with B. D\'{o}ra, A. Virosztek and S. Tomi\'{c}.

This paper is dedicated to memory of Michael J. Rice. His interest and support of our work and our friendship
over many years have been important to us.

\end{document}